\documentclass
[aps,prl,twocolumn,preprintnumbers,amssymb,nofootinbib,showpacs]{revtex4}%
\pdfoutput=1
\usepackage{graphicx}
\usepackage{dcolumn}
\usepackage{bm}
\usepackage{amssymb}
\usepackage{amsmath}
\usepackage{multirow}
\usepackage[usenames]{color}
\usepackage{hyperref}
\usepackage{amsfonts}

\providecommand{\U}[1]{\protect\rule{.1in}{.1in}}

\begin{document}
\title{Techniques for improved heavy particle searches with jet substructure}
\author{Stephen D. Ellis}
\author{Christopher K. Vermilion}
\author{Jonathan R. Walsh}
\affiliation{University of Washington, Seattle, WA 98195-1560}

\begin{abstract}
We present a generic method for improving the effectiveness of heavy particle
searches in hadronic channels at the Large Hadron Collider. By selectively
removing, or \textit{pruning}, protojets from the substructure provided by
a $\text{k}_{\text{T}}$-type jet algorithm, we improve the mass resolution
for heavy decays and
decrease the QCD background. We show that the protojets removed are typical of
soft radiation and underlying event contributions, and atypical of accurately
reconstructed heavy particles.

\end{abstract}

\maketitle


The Large Hadron Collider (LHC) presents at once great opportunity and great challenge. Many scenarios for new physics involve heavy particles that decay, possibly through a cascade, to Standard Model (SM) light quarks and gluons.  The resulting final states consist partly, or even entirely, of jets. If the new particles are not too heavy, they may often be produced with sufficient boost to appear in a single jet. Thus, in the search for new physics at the LHC, identifying those jets that contain the decay of a heavy particle may be an important tool. The key difficulty will be separating this signal from the SM background, namely QCD jets. Recently, several groups have suggested novel and effective techniques for separating hadronic decays of heavy particles from QCD making use of the expected differences in the internal structure of the jets \cite{Butterworth_cites, Brooijmans:08.1, Butterworth:08.1, Thaler:08.1, Kaplan:08.1, Almeida:08.1, Butterworth:09.1}.
The procedures proposed tend to be ``top-down'' in the sense that they are tuned to specific properties of, say, the two-pronged decay of a Higgs boson, or the three-pronged decay of a top quark.  Here we present a related approach, based, of course, on the same underlying differences between real decays and QCD, but of a simpler nature and intended for use in general searches for new (\emph{a priori} unknown) heavy particles.

While historically the masses of jets have played little role in the analysis
of collider data, this is likely to change at the LHC \cite{Ellis:08.1}. The
simplest way to search for heavy particle decays into single jets is to look
for features (``bumps'') in the jet mass
distribution for an observed jet sample. Since QCD lacks any intrinsic scale
beyond $\Lambda_{QCD}$, the background will be featureless aside from
statistical fluctuations. Further, if the heavy particle decay includes a
chain of new heavy particles, it is natural to ask whether we can look for
evidence of these other mass scales in the \emph{substructure} of the jet. Consider, for
example, searching for a top quark in a single
jet (as in \cite{Thaler:08.1, Kaplan:08.1, Almeida:08.1}). (We will use the top
quark as a surrogate for new particle searches in the studies outlined below.)
We would not only expect to see an enhancement for jet masses near the top quark
mass, but we would expect correlated evidence of the $W$ boson mass in the
substructure of the jet. If we are using a recombination algorithm such as
the $\text{k}_{\text{T}}$ algorithm, the natural choice is to identify
the $W$ with one of the protojets involved in the final merging.

Our aim in this paper is to present a procedure that improves the
effectiveness of this type of search. Our technique suppresses systematic
effects of the jet algorithm, as well as generic features of hadron collider
events, such as the underlying event. Both effects tend to obscure the mass
scales present in a heavy particle decay as observed in a single jet. Our
technique narrows the structure in the jet and protojet mass distributions for
jets from heavy particle decays, and reduces the smooth background QCD jet
mass distribution. The result is a substantially increased likelihood of
identifying a new physics (heavy particle) signal in the measured jet and
protojet mass distributions.

Jet algorithms are designed to interpret long-distance degrees of freedom
observed in the detector in terms of short-distance degrees of freedom. The
algorithms take a set of initial protojets, such as calorimeter
towers, and group them into jets. Recombination algorithms are a special class
of jet algorithms that specify a prescription to pairwise combine protojets in
an iterative procedure, eventually yielding jets. This prescription is based
on the dominant soft and collinear physics in the QCD shower, so that the
algorithm can trace back to objects coming from the hard scattering. The
pairwise merging scheme of recombination algorithms naturally gives
substructure to a jet, which provides kinematic handles to determine whether
the jet was produced by QCD alone or a heavy particle decay plus QCD.

A general recombination algorithm uses a distance measure $\rho_{ij}$ between
protojets to control how they are merged. A \emph{beam distance} $\rho_{i}$
determines when a protojet should be promoted to a jet. The algorithm
iteratively finds the smallest of the $\rho_{ij}$ and the $\rho_{i}$. If the
smallest is a $\rho_{ij}$, protojets $i$ and $j$ are merged into a new
protojet. Otherwise, the protojet corresponding to the smallest $\rho_{i}$ is
promoted to a jet. The algorithm terminates when no protojets remain.

For the $\text{k}_{\text{T}}$ \cite{kTcites} and
Cambridge-Aachen (CA) \cite{Dokshitzer:97.1} algorithms, the metrics are
\begin{equation}%
\begin{split}
\text{k}_{\text{T}}:\rho_{ij}\equiv\min(p_{T_{i}},p_{T_{j}})\frac{\Delta
R_{ij}}{D},\quad &  \quad\rho_{i}\equiv p_{T_{i}};\\
\text{CA}:\rho_{ij}\equiv\frac{\Delta R_{ij}}{D},\qquad\ \ \qquad\qquad &
\quad\rho_{i}\equiv1;
\end{split}
\end{equation}
where $p_{T_{i}}$ is the transverse momentum of protojet $i$ and $\Delta
R_{ij}\equiv\sqrt{(\phi_{i}-\phi_{j})^{2}+(y_{i}-y_{j})^{2}}$ is a measure of
the angle between two protojets, where $\phi$ is the azimuthal angle around
the beam direction and $y$ is the rapidity along the beam direction. \ The
angular parameter $D$ governs when protojets should be promoted to jets: it
determines when a protojet's beam distance is less than the distance to other
objects. The substructure arising from this pairwise merging procedure is
straightforward.

In considering the kinematics of the substructure, two variables, $z$ and
$\theta$, are particularly useful. For a recombination $1,2\rightarrow p$, we
define
\begin{equation}
z\equiv\min(p_{T_{1}},p_{T_{2}})/p_{T_{p}},\qquad\theta\equiv\Delta R_{12}.
\end{equation}
To identify heavy particle decays reconstructed in a single jet, we are
concerned with recombinations that occur at large $\theta$, typically the
final recombination. In general, small-$\theta$ recombinations are likely to
represent the QCD showering of the decay products. Similarly, small-$z$, or
soft, recombinations are typical for a QCD shower. Even the large-angle, but
small-$z$, recombinations that can appear in jets from a heavy particle decay
will be unlikely
to yield an accurate representation of the decay: if a heavy
particle decays such that one decay product has a much lower $p_{T}$ relative to the
others, the parent particle is unlikely to be accurately reconstructed. \ So,
while the variable $z$ can be an effective discriminator between QCD and
decays in principle, the substructure found by the jet algorithm often does
not faithfully represent the differing dynamics. Soft radiation, as well as
soft contributions from the underlying event and pileup, will be present in
all jets. These contributions to the jet lead to broadened mass distributions,
especially for $\text{k}_{\text{T}}$ jets. In addition, due to the systematic
effects of the jet algorithm, these soft contributions can often appear in the
final recombination. This is particularly true for CA jets, because CA
orders strictly by $\theta$.\ \ The large number of soft protojets ensures
that frequently one will appear at a large angle in the final recombination.

We now define a procedure that systematically removes these undesirable soft,
large angle recombinations.  The procedure operates by rerunning the algorithm
and vetoing on these recombinations, i.e., removing, or \emph{pruning}, them from
the substructure of the jet.  It is algorithmically similar
to others \cite{Butterworth:08.1, Kaplan:08.1}, which also modify the jet
substructure to improve heavy particle identification.
The key distinction is that pruning is applied to an entire jet
from the bottom up, with no goal of finding a particular
number of ``subjets''. The pruning procedure is:

\begin{itemize}
\item[1.] Rerun the jet algorithm on the set of initial protojets from the
original jet, checking for the following condition in each recombination
$1,2\to p$:
\begin{equation}
z < z_{\text{cut}} \qquad\text{and} \qquad\Delta R_{12} > D_{\text{cut}} .
\end{equation}

\item[2.] If this condition is met, do not merge the two protojets $1$ and $2$
into $p$. Instead, discard the softer protojet and proceed with the algorithm.
The resulting jet is the \emph{pruned} jet.
\end{itemize}

The pruning procedure involves two parameters, $z_{\text{cut}}$
and $D_{\text{cut}}$, which determine how small $z$ must be and the minimum
angle $\Delta R$ of the recombination for it to be pruned.  In this study we
use $D_{\text{cut}}=m_{{J}}/p_{{T_{J}}}$ for both $\text{k}_{\text{T}}$ and CA, where
$m_{J}$ is the mass of the originally identified jet and $p_{T_{J}}$ is its
transverse momentum.  This choice is both adaptive to the properties of the
individual jet and IR safe.
Pruning with a smaller $D_{\text{cut}}$ degrades the mass resolution by
significantly pruning the QCD shower of daughter partons of the heavy particle
decay, and pruning with a larger $D_{\text{cut}}$ does not take full advantage
of the procedure. For the CA algorithm, we use $z_{\text{cut}}=0.10$. Because
the $\text{k}_{\text{T}}$ algorithm orders recombinations partly in $z$, very
small-$z$ recombinations are not expected at the end of the algorithm. This
implies a more aggressive pruning procedure is needed for the $\text{k}%
_{\text{T}}$ algorithm, so in this study we use $z_{\text{cut}}=0.15$ for the
$\text{k}_{\text{T}}$ algorithm. We find that these values of the pruning
parameters yield roughly optimal results, largely insensitive to small changes
in their values \cite{Ellis:09.2}.

We examine the effects of the pruning procedure in a study of top quark
reconstruction and separation from the QCD background. The top quark serves as a
surrogate for a heavy particle decay at the LHC, and lets us learn about the
effects of pruning in identifying heavy particles.

We generate events using MadGraph/MadEvent v4.4.21 \cite{Alwall:07.1} interfaced
with Pythia v6.4 \cite{Sjostrand:06.1} for showering and hadronization.  For the
QCD background, we produce a matched sample of 2, 3, and 4 hard partons (gluons
and the four lightest quarks) using MLM-style matching implemented
in MadGraph (see, e.g., \cite{Alwall:08.1}).  We use the DWT tune \cite{Albrow:06.1} in
Pythia to give a ``noisy'' underlying event. No detector
simulation is performed so we can isolate the ``best case'' effects of our
method.

The signal sample is $t\bar{t}$ production with fully hadronic decays.  We generate
signal and background samples with a parton-level $h_{T}$ cut for
generation efficiency, where $h_{T}$ is the scalar sum of all $p_{T}$ in the event.
Because we focus on single jet methods to identify
heavy particles, we make samples defined by criteria on jets instead of
events. For each sample, we select central jets (with pseudorapidity $|\eta| < 2.5$) and divide them
into four $p_T$ bins: [200, 500], [500, 700], [700, 900], and [900, 1100] (all in GeV/c).
These bins confine the top quark boost to a narrow range within each bin and allow us to
study the performance of pruning as the top quark $p_{T}$ varies.  For
each $p_T$ bin $[p_T^\text{min}, p_T^\text{max}]$, the parton-level $h_T$ cut
is $p_T^\text{min} - 25\ \text{GeV/c} \le h_T/2 \le  p_T^\text{max} + 100\ \text{GeV/c}$.
We take the matching
scales $(Q^\text{ME}_\text{cut}, Q_\text{match})$ to be (20, 30) GeV for the lowest $p_T$ bin
and (50, 70) GeV in the other three bins.

From the hadron-level output of Pythia, we group final-state particles into
``cells'' based on the segmentation of the
ATLAS hadronic calorimeter ($\Delta \eta=0.1$, $\Delta\phi=0.1$ in the central region). We sum the
four-momenta of all particles in each cell and rescale the resulting
three-momentum to make the cell massless. After a threshold cut on the cell
energy of 1 GeV, cells become the inputs to the jet algorithm. Our implementation of
recombination algorithms uses FastJet \cite{Cacciari:05.1}.

\begin{figure}[th]
\begin{center}
\includegraphics[width=8.5cm]{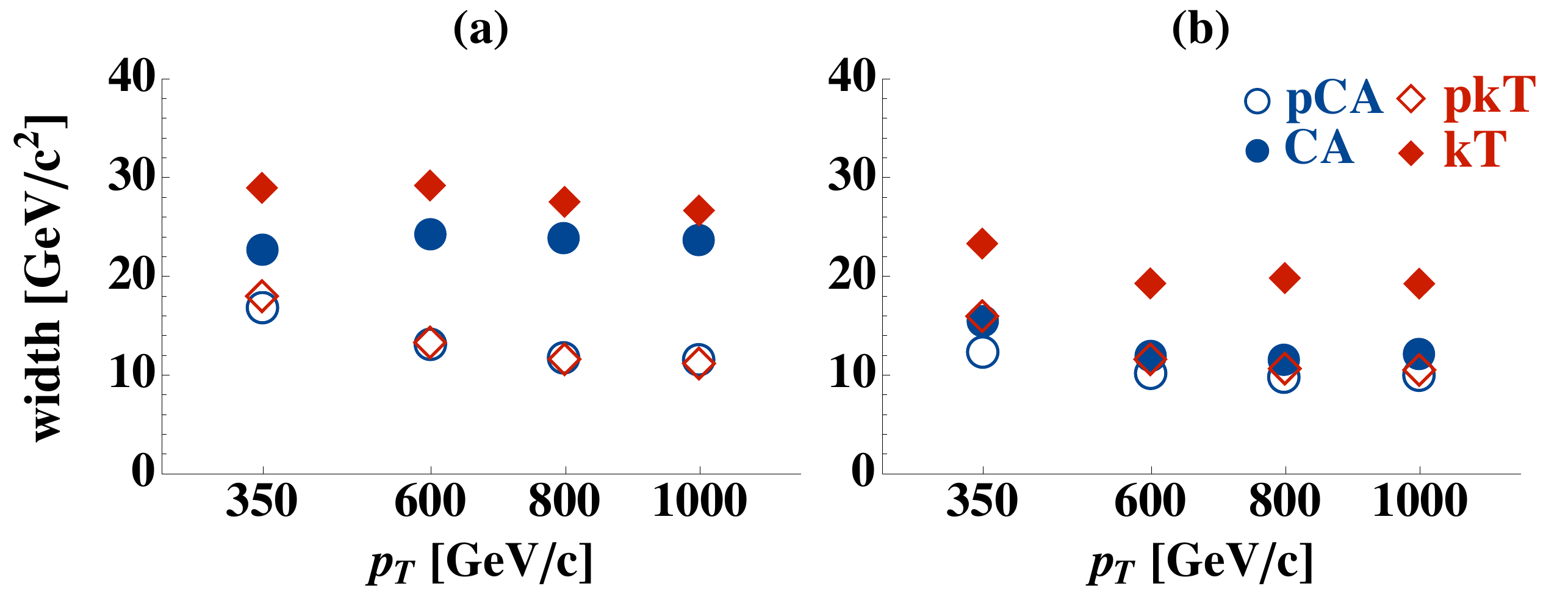}
\end{center}
\caption{Pruned and unpruned top (a) and $W$ (b) mass window widths
(in GeV/$\text{c}^2$) versus $p_{T}$ window center
(in GeV/c) for both $\text{k}_{\text{T}}$ and CA algorithms.}%
\label{windowwidths}%
\end{figure}

To quantify the effects of pruning in top identification and background
separation, we define criteria for a jet to be labeled as reconstructing a top
quark decay. For either the pruned or unpruned jet, a \emph{top jet} is one
whose mass is within the top mass window and whose heavier daughter protojet mass is
within the $W$ mass window. Both windows come from fits to the mass
distributions in the signal sample, and do not need to be known \emph{a
priori}. These are fit using a skewed Breit-Wigner distribution for the peak
and a power-law continuum for the background. These functions are
\begin{equation}%
\begin{split}
\text{peak: }f(m) &  =M^{2}\Gamma^{2}\frac{\left[  a+b(m-M)\right]  }%
{(m^{2}-M^{2})^{2}+M^{2}\Gamma^{2}};\\
\text{continuum: }g(m) &  =\frac{c}{m}+\frac{d}{m^{2}}.
\end{split}
\end{equation}
The fitted mass $M$, which is within a few GeV/$\text{c}^2$ of
$m_{\text{top}}$, and the
fitted width $\Gamma$ are the relevant parameters; the mass window is
$M\pm\Gamma$. These mass windows are in general different for the pruned and
unpruned samples. In Fig.~\ref{windowwidths}, we plot the top and $W$ window
widths for the $\text{k}_{\text{T}}$ and CA algorithms for both pruned and
unpruned jets. We refer to the pruned version of algorithm $A$ as $pA$.

The top and $W$ mass windows are significantly narrower for the pruned
samples. Moreover, the widths for the pruned $\text{k}_{\text{T}}$ and CA
algorithms are very similar, unlike the unpruned case. The narrower widths
mean fewer jets from the QCD samples will be misidentified as tops.

\begin{figure}[th]
\begin{center}
\includegraphics[width=8.5cm]{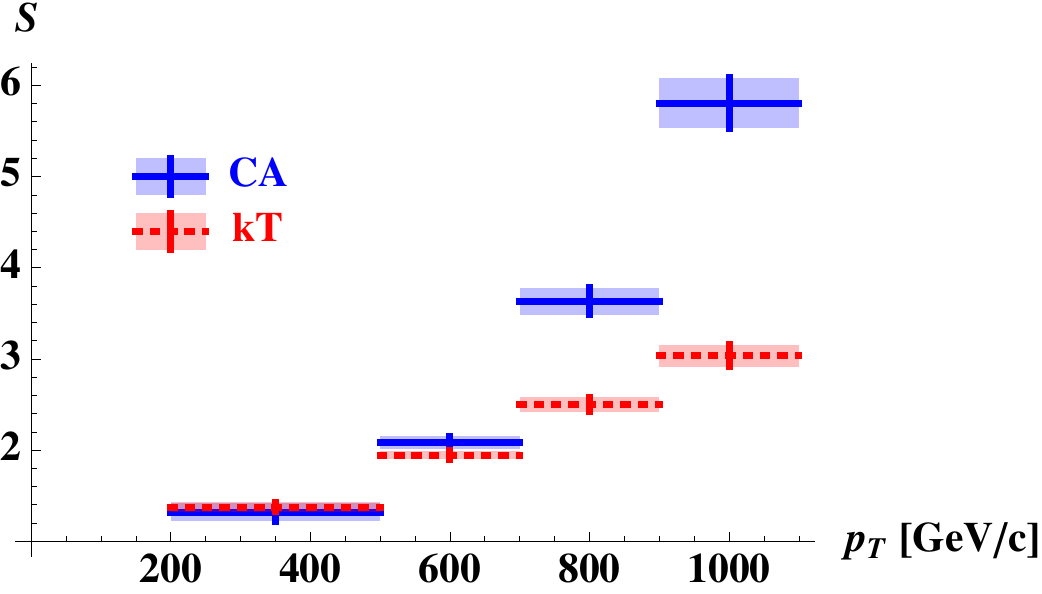}
\end{center}
\caption{$S$ vs. $p_{T}$ for the CA and $\text{k}_{\text{T}}$
algorithms, using $D = 1.0$. Statistical errors, due to limited QCD sample sizes after cuts, are shown.}%
\label{SvarypT}%
\end{figure}

We now discuss a more quantitative measure of the performance of pruning. From
the found mass windows we count the number of top jets in the signal and
background samples, $N_{\text{{\tiny {S}}}}(A)$ and $N_{\text{{\tiny {B}}}%
}(A)$, for algorithm $A$. Using these counters, we define a statistical
measure, $S$, to quantify how pruning improves top identification and
separation from QCD backgrounds. $S$ is defined as
\begin{equation}
S=\frac{N_{\text{{\tiny {S}}}}(pA)/\sqrt{N_{\text{{\tiny {B}}}}(pA)}%
}{N_{\text{{\tiny {S}}}}(A)/\sqrt{N_{\text{{\tiny {B}}}}(A)}},
\end{equation}
which is the improvement from pruning in the ratio of the signal size to the
statistical fluctuations in the background, and is a measure of the expected
improvement in significance of the signal. Values greater than one indicate an
improvement in pruning versus not pruning. Note that while the significance of the
signal depends on the relevant cross sections and the integrated luminosity,
the improvement measure $S$
does not. Using a constant value of $D=1.0$ for all $p_{T}$ bins, we
plot $S$ in Fig.~\ref{SvarypT} for both the $\text{k}_{\text{T}}$ and CA algorithms.

For both algorithms, the measure $S$ is in the range 1.2 to 1.4 in
the lowest $p_{T}$ bin, and increases with increasing $p_{T}$, with a
dramatically increased significance in the range of 3 to 6 in the highest bin. These
large values of $S$ arise partially from using a fixed value of $D$ with
varying $p_{T}$. The opening angle of the typical top quark decay varies as
$\Delta R \approx2m_{\text{top}}/p_{T}$, which is less than $D=1.0$ in the larger $p_{T}$ bins.
The large $D$ allows for extra radiation to be merged in the
jet, which may sufficiently alter the order of the substructure reconstruction
to render an actual top decay
no longer identifiable as a top jet. Additionally, a larger $D$ at fixed
$p_T$ leads to larger mass QCD jets and enhances the probability to
fake top quarks. In both scenarios the extra radiation included within
the larger $D$ jet is often soft and uncorrelated.  Hence pruning tends to dramatically
improve top finding at large $p_T$ in fixed $D$ jets.

In a real search, the mass of the heavy state is not known. Once an
enhancement in
the mass distribution has been observed, knowledge of the purported mass can
be used to tune the analysis parameters, such as $D$.
(Another approach, using ``variable-R'' jets, is discussed in \cite{Krohn:09.1}.)
Even if $D$ is tuned
for each $p_{T}$ bin to maximize the performance of the unpruned algorithm, we
would still expect pruning to show an improvement over the unpruned case. This
can be seen in the lowest bin of Fig.~\ref{SvarypT}, where the value of $D = 1.0$ is
already roughly optimal and $S$ is still larger than 1.

\begin{figure}[th]
\begin{center}
\includegraphics[width=8.5cm]{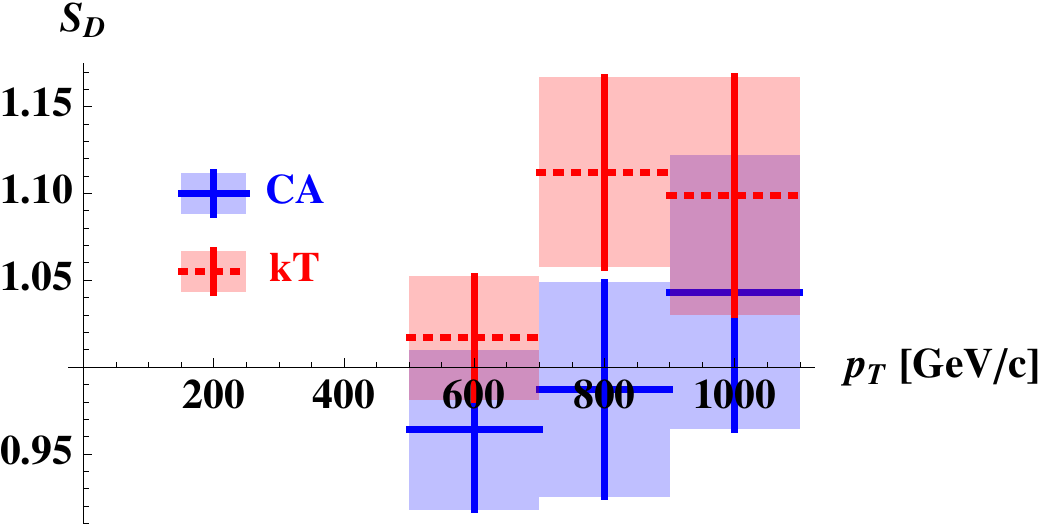}
\end{center}
\caption{$S_{D}$ vs. $p_{T}$ for the CA and $\text{k}_{\text{T}}$
algorithms. The line at $S_{D}$ = 1 separates the regions
where a tuned $D$ helps (above the line) and does not (below). The lowest
$p_{T}$ bin is not shown because the $D$ value does not change ($S_{D} =
1$).  Statistical errors are shown.}%
\label{SDpruning}%
\end{figure}

Given that pruning always
provides an improvement, the relevant question for designing a search procedure
using single jets is whether pruned, tuned-$D$ jets provide much better
results than pruned, fixed-$D$ jets.  To answer this question, we
compare signal-to-noise for pruned jets with fixed $D = 1.0$ to the case
where $D$ is picked for each
$p_{T}$ bin to match the typical opening angle of the top quark decay. In
particular, we set $D$ to be approximately $2m_{\text{top}}/p_{T}^{\min}$,
where $p_{T}^{\min}$ is the lower $p_{T}$ limit for the given bin, up to a
maximum of 1.0. Thus we choose the $D$ values of $\{1.0,0.7,0.5,0.4\}$ for
our $p_T$ bins.  This exercise leads to Fig.~\ref{SDpruning},
where we show a ratio analogous to $S$ that we call
$S_{D}$. For each $p_{T}$ bin, $S_{D}$ is the ratio of
signal-to-noise for pruned jets with the value of $D$ from the above list
to signal-to-noise for pruned jets with fixed $D=1.0$. We see that the
values of $S_{D}$ are close to one for all
$p_{T}$ bins. This implies the important result that, as long as we prune
the jets, using a tuned $D$ value for each $p_{T}$ bin provides little advantage
over the simpler fixed
$D$ analysis. Note also that in Fig.~\ref{SDpruning}
the statistical uncertainties in $S_{D}$ are on
the order of the improvements.

In this work, we have introduced a generic procedure that modifies jet
substructure to improve heavy particle identification and separation from QCD
backgrounds. This procedure, pruning, removes recombinations unlikely to
represent an accurately reconstructed heavy particle, narrows mass
distributions of reconstructed states and reduces the QCD background in a
given mass bin.  As we have demonstrated, heavy particle searches can benefit
from all of these effects. While unpruned jets are sensitive to the
specific choice of jet algorithm and the value of the parameter
$D$, pruning
removes much of this sensitivity.  It is just as effective to use a large
$D$ over a broad range in $m/p_{T}$ of the heavy state. When searching for a
particle of unknown mass, pruning allows the use of a large fixed $D$ without
losing statistical power.

The effects of pruning, and in general the application of jet substructure
to find heavy particles, requires further study \cite{Ellis:09.2}. Pruning
must be verified as an effective component of heavy particle searches at the
LHC, including understanding the impact of using a realistic detector.  An
important test bed for pruning and other jet substructure tools will be early
validation studies of the Standard Model at the LHC, where we expect to be
able to observe top quarks, $W$'s and $Z$'s in the single jet data. Initial
studies such as that described here give promising indications that these
tools will prove useful in the search for new physics.

We gratefully acknowledge helpful discussions with Matt Strassler and Gavin Salam,
and also
with Karl Jacobs, Peter Loch, Michael Peskin, Tilman Plehn, and others in the
context of the Joint Theoretical-Experimental Terascale Workshops at the
Universities of Washington and Oregon, supported by the U.S. Department of
Energy under Task TeV of Grant No.~DE-FG02-96ER40956. This work was supported in
part by the U.S.
Department of Energy under Grant No.~DE-FG02-96ER40956. J.W. was also
supported in part by an LHC Theory Initiative Graduate Fellowship.

\bibliography{jetcites}

\end{document}